\documentclass[aps,prl,english,reprint,footinbib,longbibliography]{revtex4-1}
\usepackage[T1]{fontenc}
\usepackage[mathletters]{ucs}
\usepackage[utf8x]{inputenc}
\usepackage{textcomp}
\usepackage{mathrsfs}
\usepackage{mathtools}
\usepackage{amsmath}
\usepackage{amssymb}
\usepackage{graphicx}
\usepackage{babel}
\usepackage{color}
\usepackage{siunitx}
\usepackage[normalem]{ulem}
\usepackage{physics}
\usepackage{xcolor}
\usepackage{bm}
\usepackage{ulem}

\begin{document}
\newcommand*\absq[1]{\abs{#1}^2}
\newcommand*\conj[1]{\overline{#1}}
\newcommand*\mred[0]{m_\text{red}}
\newcommand*\aBB[0]{a_\text{BB}}
\newcommand*\aIB[0]{a_\text{IB}}
\newcommand*\VIB[0]{V^\text{IB}}
\newcommand*\VBB[0]{V^\text{BB}}
\newcommand*\mI[0]{m_\text{I}}
\newcommand*\mB[0]{m_\text{B}}
\newcommand*\sigmaBB[0]{\sigma_\text{BB}}
\renewcommand*\vec[1]{\bm{#1}}

\allowdisplaybreaks

\title{Theory of a resonantly interacting impurity in a Bose-Einstein condensate}
\author{Moritz Drescher}
\author{Manfred Salmhofer}
\author{Tilman Enss}
\affiliation{Institut für Theoretische Physik, Universität Heidelberg,
  D-69120 Heidelberg, Germany}

\begin{abstract}
  We investigate a Bose-Einstein condensate in strong interaction with
  a single impurity particle.  While this situation has received
  considerable interest in recent years, the regime of strong coupling
  remained inaccessible to most approaches due to an instability in
  Bogoliubov theory arising near the resonance.  We present a nonlocal
  extension of Gross-Pitaevskii theory that is free of such
  divergences and does not require the use of the Born approximation
  in any of the interaction potentials.  We find a new dynamical
  transition regime between attractive and repulsive polarons, where
  an interaction quench results in a finite number of coherent
  oscillations in the density profiles of the medium and in the
  contact parameter before equilibrium is reached.
\end{abstract}

\maketitle

The motion of an impurity through a dynamical medium consisting of
many interacting particles or field excitations is a fundamental
physical problem.  The impurity gets dressed by a cloud of excitations
of the medium, which may enhance its mass (or even generate it in the
first place, as in the Higgs effect) and also renormalize other
properties \cite{Girardeau1961, Devreese2009}.  Conversely, the medium
is deformed in the vicinity of the impurity.  This blurs the
distinction between impurity and medium, and new many-body excitations
and entangled states emerge \cite{Boyanovsky2019}.  In particular, an
impurity subject to a strong attractive interaction with the
particles in the medium may form bound states with them.  Such
many-body bound states allow one to probe the properties of the medium
itself and occur, e.g., in ultracold gases \cite{Rath2013,
  Grusdt2017a}, Rydberg systems \cite{Camargo2018}, semiconductor
heterostructures \cite{Sidler2017}, or as dissipative bound states in
photonic crystals and matter waves \cite{Liu2017, Krinner2018}.

Recent experiments with ultracold quantum gases have probed impurity
atoms in a Bose-Einstein Condensate (BEC) medium, so-called Bose
polarons \cite{Catani2012, Rentrop2016, Hu2016, Jorgensen2016,
  Camargo2018, PenaArdila2019, Yan2020}.  In these systems, the
impurity-boson (IB) interaction can be tuned by a Feshbach scattering
resonance to strong attraction and even to the formation of two-body
bound states between impurity and bosons \cite{Chin2010}.  While such
bound states are well understood on the the two-body level, they lead
to a new competition on the many-body level: the short-range
attraction tends to bind as many medium particles as possible to the
impurity and strongly enhances the local medium density, while the
short-range boson-boson (BB) repulsion between the medium particles
counteracts this effect and tends to spread the bosons evenly.  The
competition between impurity-boson and boson-boson interaction can
cause new dynamical many-body effects such as quantum flutter
\cite{Mathy2012}.

Previous theoretical approaches based on the Fr\"ohlich Hamiltonian
\cite{Frohlich1954, Girardeau1961, Astrakharchik2004, Cucchietti2006,
  Bruderer2008, Tempere2009, Casteels2011, Shashi2014, Grusdt2015,
  Vlietinck2015, Lampo2017, nielsen2019, Boyanovsky2019} accurately
describe the three-dimensional (3D) Bose polaron at weak coupling.  In
order to capture the bound state at strong coupling, it is necessary
to go beyond the Fr\"ohlich model and include quadratic coupling terms
\cite{Rath2013, Li2014, Christensen2015, Levinsen2015, PenaArdila2015,
  Volosniev2015, Shchadilova2016, Grusdt2017a, Grusdt2018, Kain2018,
  Lausch2018, Drescher2019, Mistakidis2019}.  Even then, the use of
the Bogoliubov approximation misses cubic and quartic interaction
terms, which are a crucial part of the boson repulsion required to
balance the local attraction and prevent an unlimited growth of local
density, as pointed out in \cite{Grusdt2017a, Drescher2019}.
Alternatively, Gross-Pitaevskii theory fully incorporates boson
repulsion but does not describe bound states with a local interaction
in Born approximation \cite{Massignan2005, Blinova2013,
  Takahashi2019}.

In this work, we develop a \emph{nonlocal} extension of
Gross-Pitaevskii theory (GPT) that captures both competing effects on
equal footing.  This requires incorporating the explicit spatial form
of the competing short-range attractive and repulsive interaction
potentials.  We apply our approach to study quench dynamics resulting
from suddenly switching on the interaction between impurity and
bosons: for this case, we predict stable and long-lived coherent
oscillations of the medium density and correlations in space and time
on the repulsive side of a scattering resonance; these are within
reach of experimental observation.  Near the (phase) boundary between
the attractive and repulsive Bose polaron dynamics close to resonance,
we find a new dynamical transition regime where a finite number of
coherent oscillations occur before an attractive polaron is formed.

\textit{Model.---}%
The Hamiltonian of a Bose gas with a single impurity is, in first
quantized form,
\begin{align*}
  H ={}& \frac{\hat{\vec p}_\text{I}^2}{2\mI} + ∑_i \frac{\hat{\vec p}_{\text{B},i}^2}{2\mB} \\
       & + ∑_i \VIB(\hat{\vec x}_{\text{B},i} - \hat{\vec x}_\text{I})
         + ∑_{i<j} \VBB(\hat{\vec x}_{\text{B},i} - \hat{\vec x}_{\text{B},j}) \, .
\end{align*}
where indices $I$ and $B$ stand for impurity and bosons with masses
$\mI$ and $\mB$, respectively, and $i, j$ enumerate the bosons.  We
assume a repulsive BB interaction $\VBB(\vec x)>0$ and an attractive
IB interaction $\VIB(\vec x)<0$.

For a single impurity it is convenient to use relative coordinates
comoving with the impurity, making use of total momentum conservation.
This is achieved by the Lee-Low-Pines (LLP) transformation
\cite{Lee1953, Girardeau1961} generated by
$S = \hat{\vec x}_\text{I} ⋅ ∑_i \hat{\vec p}_{\text{B},i}$.  Writing
$\vec p_0$ for the conserved total momentum and $\vec x_i$ for the
relative coordinate of the $i^\text{th}$ boson, we obtain the LLP
Hamiltonian $H_\text{LLP} = \exp(iS) H \exp(-iS)$:
\begin{align} \label{eq:LLP Hamiltonian}
  H_\text{LLP}
  ={}& \frac{\qty(\vec p_0 - ∑_i \hat{\vec p}_i)^2}{2\mI}
       + ∑_i \frac{\hat{\vec p}_i^2}{2\mB} \nonumber \\
     & + ∑_i \VIB(\hat{\vec x}_i) + ∑_{i<j} \VBB(\hat{\vec x}_i -
       \hat{\vec x}_j) \, .
\end{align}

\textit{Gross-Pitaevskii theory in relative coordinates.---}%
The case of a long-range impurity potential $\VIB$ can be described by
standard GPT \cite{Pitaevskii2003}.  In particular, this includes
interesting cases where the impurity potential has a bound state,
which mimics the repulsive side of a Feshbach resonance.  When this
potential is suddenly switched on, long-lived coherent oscillations
appear from the interference between the bound state and the
homogeneous BEC state \cite{Drescher2019}.  In the following, we first
introduce GPT and discuss under which conditions it applies to the
Hamiltonian \eqref{eq:LLP Hamiltonian}; we then develop the new theory
that works also for competing short-range potentials.

A heuristic derivation of GPT consists in replacing a radial BB
potential by a local term in Born approximation,
$\VBB(r) → (4π \aBB/\mB) δ^{(3)}(r),$ where $\aBB$ is the scattering
length, and making a product ansatz
\begin{equation}
  \label{eq:product}
  Ψ(\vec x_1, …, \vec x_N) = ϕ(\vec x_1) ⋯ ϕ(\vec x_N) \, ,
\end{equation}
where $ϕ$ is normalized as $∫ d^dx\, \absq{ϕ(\vec x)} = N$.

In the thermodynamic limit one derives the energy functional
\begin{align} \label{eq:GP functional}
  E_\text{GP}[ϕ]
  ={}& ∫_{\vec x} \qty( \frac {\absq{∇ϕ}} {2\mred} + \VIB \absq{ϕ}
              + \frac{4π \aBB}{2\mB} \abs{ϕ}^4) \nonumber \\
     & + \frac {\qty(\vec p_0 - ∫ \conj{ϕ} \hat{\vec p} ϕ)^2} {2\mI}
\end{align}
with reduced mass $\mred^{-1} = \mI^{-1} + \mB^{-1}$.  It differs from
the usual GP energy functional in three ways: $(i)$ the external
potential is replaced by the impurity potential, which is static in
the LLP frame; $(ii)$ the kinetic term contains the reduced mass
instead of the boson mass, and $(iii)$ the term in the second line
induces correlations between bosons due to the motion of the
impurity. It is non-constant only when $ϕ$ is not spherically
symmetric: within the product ansatz \eqref{eq:product}, this is the
case only when the impurity is initially moving with respect to the
condensate ($\vec p_0 ≠ 0$).

\textit{Applicability of GPT to Bose polarons.---}%
GPT is usually applied to a Bose gas confined in an external potential
that varies on a length scale larger than the range of the BB
potential. This separation of scales justifies the Born approximation
above.  For the system at hand, this means that the energy functional
\eqref{eq:GP functional} can only be applied when the range of the
impurity potential $\VIB$ is large, but not for arbitrary potential
shapes where the length scales no longer separate.  In particular, a
contact impurity interaction would require the boson field $ϕ(\vec x)$
to diverge as $1/\abs{\vec x}$ near the impurity $\abs{\vec x}\to0$,
but then the BB term $∫_{\vec x} \abs{ϕ(\vec x)}^4 \sim 1/r_0$
diverges as the inverse of the potential range $r_0$.  Hence, standard
GPT fails to describe competing short-range interactions as they occur
in ultracold quantum gases.

If both potentials are weak, the Born approximation may be applied to
$\VIB$ as well, hence it works when the two interaction length scales
are of the same order \cite{Blinova2013, Takahashi2019}.  Here,
however, we are interested in large IB coupling and need to look for a
theory that takes into account the interaction potential shapes in
more detail.

\textit{Generalization to strong coupling.---}%
GPT was rigorously proven to be exact for the ground state of a dilute
gas, for a large class of repulsive BB potentials and without
requiring the Born approximation \cite{Lieb2000}. An essential insight
is that the wave function is not described by a pure product state
(which would immediately lead to infinite energy in the case of a
hard-sphere potential), but rather by \cite{Dyson1957, Lieb2000}
\begin{equation} \label{eq:ansatz}
  Ψ(\vec x_1, …, \vec x_N)
  = ϕ(\vec x_1) ⋯ ϕ(\vec x_N) \; F(\vec x_1, …, \vec x_N) \, .
\end{equation}
Here, $F$ encodes two-body effects: $F=1$ when all particles are far
apart from each other, but it behaves as the zero-energy scattering
solution when any two particles are close to each other, and as a
product of such two-body wave functions when multiple two-clusters
occur.  For a soft potential, $F≈1$ everywhere and the product state
is recovered.  Specifically, we choose
\begin{equation*}
  F(\vec x_1, …, \vec x_N) = ∏_{i<j} f(\vec x_i - \vec x_j)
\end{equation*}
where $f$ is the zero-energy scattering solution of the two-boson
problem, normalized as $\lim_{|\vec x| \to \infty} f(\vec x) = 1$,
\begin{equation*}
  \qty(-\frac{∇^2}{\mB} + \VBB(\vec x)) f(\vec x) = 0 \, .
\end{equation*}
For instance, a hard-sphere potential of range $\aBB$ has
$f(\vec x) = \operatorname{max}(0, 1- \aBB / \abs{\vec x})$.  The
two-body correlated wave functions \eqref{eq:ansatz} were first
employed by Jastrow \cite{Jastrow1955}, and ansatzes of similar type
are used in rigorous studies of BEC dynamics \cite{Erdos2007,
  Erdos2009, Brennecke2019} and in quantum Monte Carlo calculations.  Taking the
expectation value of the Hamiltonian \eqref{eq:LLP Hamiltonian} in a
correlated wave function of the form \eqref{eq:ansatz}, we obtain the
energy functional $E[\phi]=\expval{H}{Ψ}/\braket{Ψ}$ in the
low-density limit as the first main result of this work,
\begin{align} \label{eq:energy functional}
  E[ϕ]
  ={}&\frac {\qty({\vec p}_0 - ∫ \conj{ϕ} \hat{\vec p} ϕ)^2} {2\mI}
  +  ∫_{\vec x} \qty( \frac {\absq{∇ϕ}} {2\mred} + \VIB \absq{ϕ} ) \nonumber \\
     & + ∫_{\vec x_1,\vec x_2} \absq{ϕ_1ϕ_2}
        \qty[\frac{f'^2_{12}} {2\mB} + \frac{\VBB_{12}} {2} \, f^2_{12}] \nonumber \\
     & + \Re ∫_{\vec x_1,\vec x_2}
     \frac{\conj{ϕ_1} ∇ ϕ_1 ⋅ \conj{ϕ_2} ∇ ϕ_2 \, \qty(1 - f_{12}^2)}
          {2\mI}\,
\end{align}
with $ϕ_1 = ϕ(\vec x_1)$, $f_{12} = f(\vec x_1 - \vec x_2)$ etc., and
$f'$ is the derivative of $f$ (since $\VBB$ is radially symmetric, $f$
only depends on $\abs{\vec x}$).  The most important difference to the
original GP functional \eqref{eq:GP functional} is the BB term (second
line), which is now a double integral and involves not only the
scattering length but the full BB potential.  The two integrals
decouple only for weak IB coupling where $ϕ_1 ≈ ϕ_2$ and the square
brackets yield the Born approximation $2π \aBB/\mB$ \cite{Lieb2000}.
At strong coupling, instead, the pair correlations contained in
$f(\vec x)$ are crucial to render the short-range interaction finite.
The last line captures the back-reaction of the medium onto the
impurity, which arises from correlations beyond the product state and
is nonzero also for an impurity initially at rest ($\vec p_0=0$).
Few-body correlations are predicted to have a strong effect on
attractive Bose polarons and yield universal properties when both
$\VIB$ and $\VBB$ are short ranged \cite{Yoshida2018,
  Shi2018impurity}.

\textit{Dynamics.---}%
\begin{figure}
  \includegraphics[width=1.0\columnwidth]{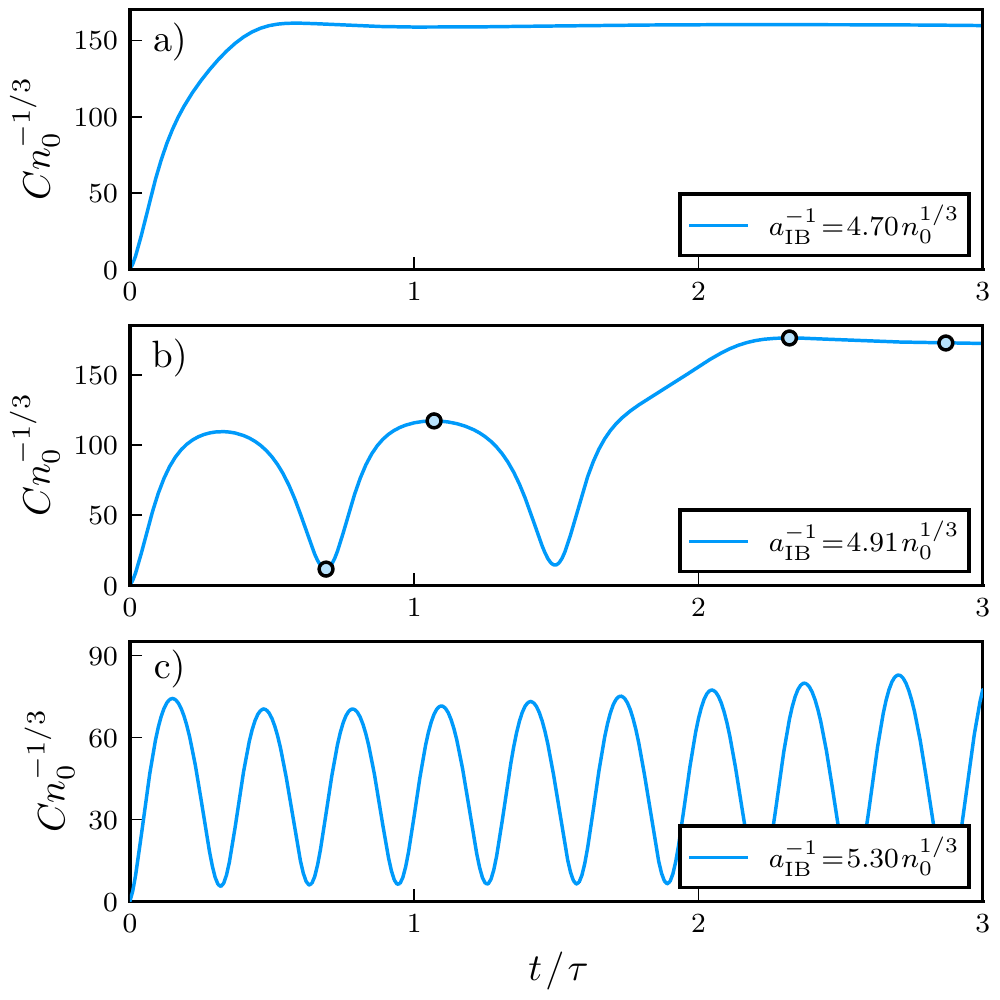}
  \caption{The contact parameter $C$ as a function of time for three
    different IB interaction strengths. (a) No oscillations are
    observed despite the positive scattering length; the polaron is
    effectively attractive due to the medium shift of the resonance in
    presence of BB interactions. (b) Close to the transition point,
    the system performs a limited number of oscillations before
    converging, featuring a dynamical transition from a
    repulsive to
    an attractive polaron. (c) Further on the repulsive side, the
    system keeps oscillating in agreement with previous predictions
    \cite{Drescher2019}.  Calculations were carried out at
    $\aBB = 0.03 n_0^{-1/3}$, $\sigmaBB = 0.1 n_0^{-1/3}$ and
    $\mB = \mI$. Time is measured in units of the BEC time scale
    $τ = \mB n_0^{-2/3}/\hbar$.}
  \label{fig:contact}
\end{figure}
We apply \eqref{eq:energy functional} to the situation where an
impurity is suddenly quenched from a noninteracting to an interacting
state, as can be realized by a hyperfine transition in experiment.  We
assume that the wave function retains the correlated form
\eqref{eq:ansatz} with $f(\vec x)$ fixed in time because the
boson-boson pair correlation is always present in the condensate, also
far away from the impurity.  In \eqref{eq:ansatz}, the time dependence
of the impurity-boson correlation is captured by $\phi(\vec x,t)$.
This is analogous to the derivation of the dynamical GP equations from
the GP energy functional; the corresponding nonlinear Schrödinger
equation is obtained by varying
$\mathcal{L} = \expval{i∂_t - H}{Ψ} / \braket{Ψ}$ with respect to
$ϕ$. This yields
\begin{align} \label{eq:time evolution}
  i\partial_t \phi_1
  ={}& -\frac{∇^2 ϕ_1}{2\mred} + \VIB_1 ϕ_1 \nonumber\\
     & + ϕ_1 ∫_{\vec x_2} \absq{ϕ_2}
       \qty[\frac{f'^2_{12}} {\mB} + \VBB_{12} f^2_{12}] \nonumber\\
     & + \frac{i ∇ ϕ_1}{\mI} ⋅ \qty(\vec p_0 -
       \Im ∫_{\vec x_2} \conj{ϕ_2} ∇ ϕ_2 \, f_{12}^2) \nonumber\\
     & + \frac{ϕ_1}{\mI} ∫_{\vec x_2} ϕ_2 ∇ \conj{ϕ_2} ⋅
       f_{12} f'_{12} \frac{\vec x_2 - \vec x_1}{\abs{\vec x_2 - \vec x_1}} \, .
\end{align}
\begin{figure}
  \includegraphics[width=1.0\columnwidth]{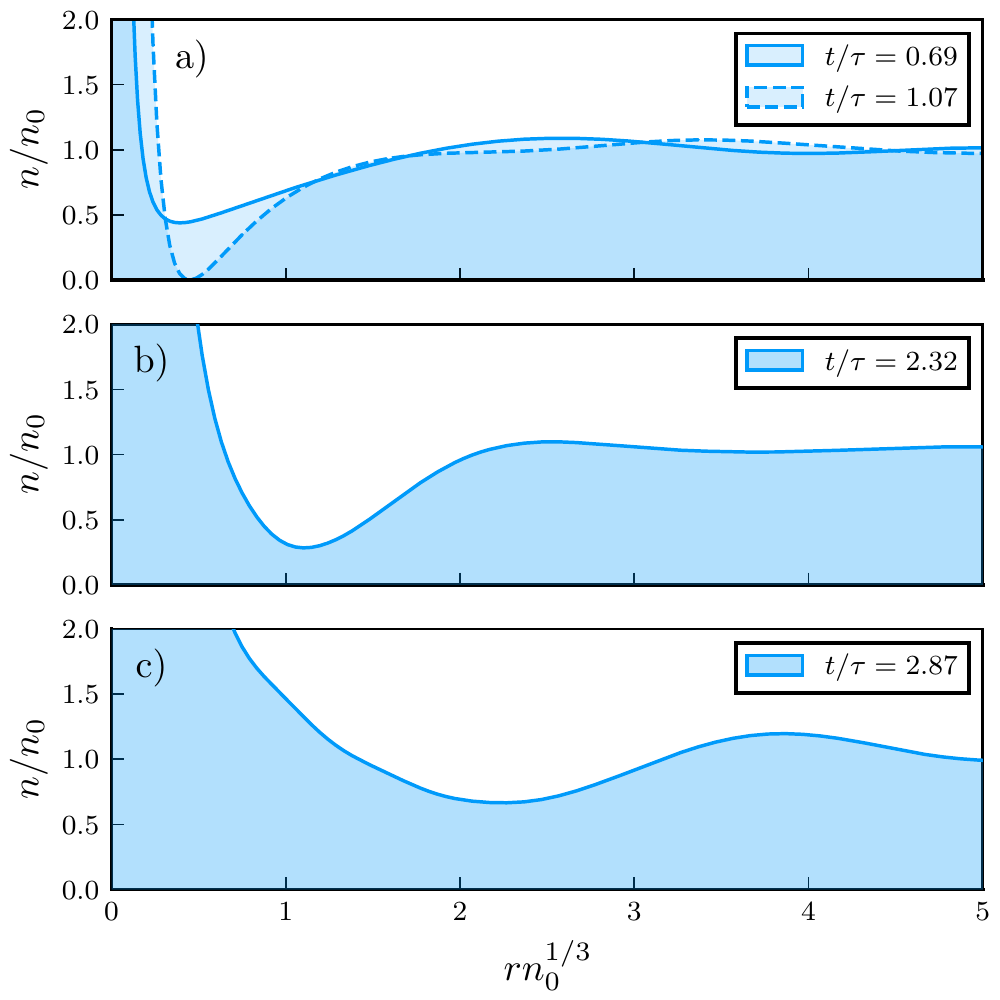}
  \caption{BEC density profiles at different times in the transition
    regime. The points in time and parameters are those indicated in
    Fig.~\ref{fig:contact}(b), while $r$ denotes the distance from the
    impurity. One can see the dynamical transition from a repulsive to
    an attractive polaron: At short times (a), a halo of depletion
    around the impurity forms and oscillates in depth
    \cite{Drescher2019}.  At intermediate times, the depletion hole
    moves outwards and becomes more shallow (b).  Finally, the profile
    of an attractive polaron is approached (c).}
  \label{fig:density}
\end{figure}

We solve this equation numerically with the initial condition of a
flat BEC, $ϕ(\vec x, t=0) = \sqrt{n_0}$, and an impurity at rest,
$\vec p_0 = 0$.  For the interaction potentials, we take $\VBB$ to be
either a Gaussian potential of depth $V_0$ and range $\sigmaBB$,
\begin{equation*}
  \VBB(r) = V_0 \exp(-r^2 / \sigmaBB^2),
\end{equation*}
or a hard-sphere potential, while $\VIB$ is modelled as a contact
interaction using the 3D Fermi pseudo-potential with scattering length
$\aIB$,
\begin{equation*}
  \VIB(r) ϕ(r) = \frac{2π \aIB}{\mred} δ^{(3)}(r) \frac{∂}{∂ r} rϕ(r) \,.
\end{equation*}
This corresponds to the Bethe-Peierls boundary condition for
$t>0$,
\begin{equation*}\label{eq:contact boundary condition}
  \lim_{r→0} r ϕ(r) + \aIB ∂_r rϕ(r) = 0 \; .
\end{equation*}
The 3D contact potential is always attractive and has no bound state
for weak attraction where $\aIB<0$ (attractive side of Feshbach
resonance).  For strong attraction beyond the scattering resonance
$1/\aIB=0$, there is a single bound state of energy
$E_B=\qty(2\mred \aIB^2)^{-1}$ and the scattering length $\aIB>0$ is
positive (repulsive side) \cite{Chin2010}.

\textit{Results.---}%
After an interaction quench, the boson density profile exhibits two
distinct phenomenologies: $(i)$ a density profile enhanced with
respect to $n_0$ that converges for long times (``attractive
polaron''), and $(ii)$ a profile depleted in a ``halo'' around the
impurity that oscillates in time (``repulsive polaron'').  Previous
approaches based on Bogoliubov theory \cite{Shchadilova2016,
  Drescher2019} have shown that quenches to the attractive side of the
Feshbach resonance yield attractive polarons, while repulsive polarons
form further on the repulsive side of the resonance; however, near the
Feshbach resonance a dynamical instability arises.  The present
approach is free of such instabilities, which allows us to investigate
the transition region near the resonance.  We find that the transition
between the many-body attractive and repulsive polaron states is
shifted toward the repulsive side of the scattering resonance due to
the many-body environment \cite{Shchadilova2016, Drescher2019}.  Our
results for equal mass $\mB=\mI$ are representative of the mass
imbalanced case within the general theory \eqref{eq:energy
  functional}.

\textit{Transition from convergent to oscillatory dynamics.---}%
\begin{figure}
  \includegraphics[width=1.0\columnwidth]{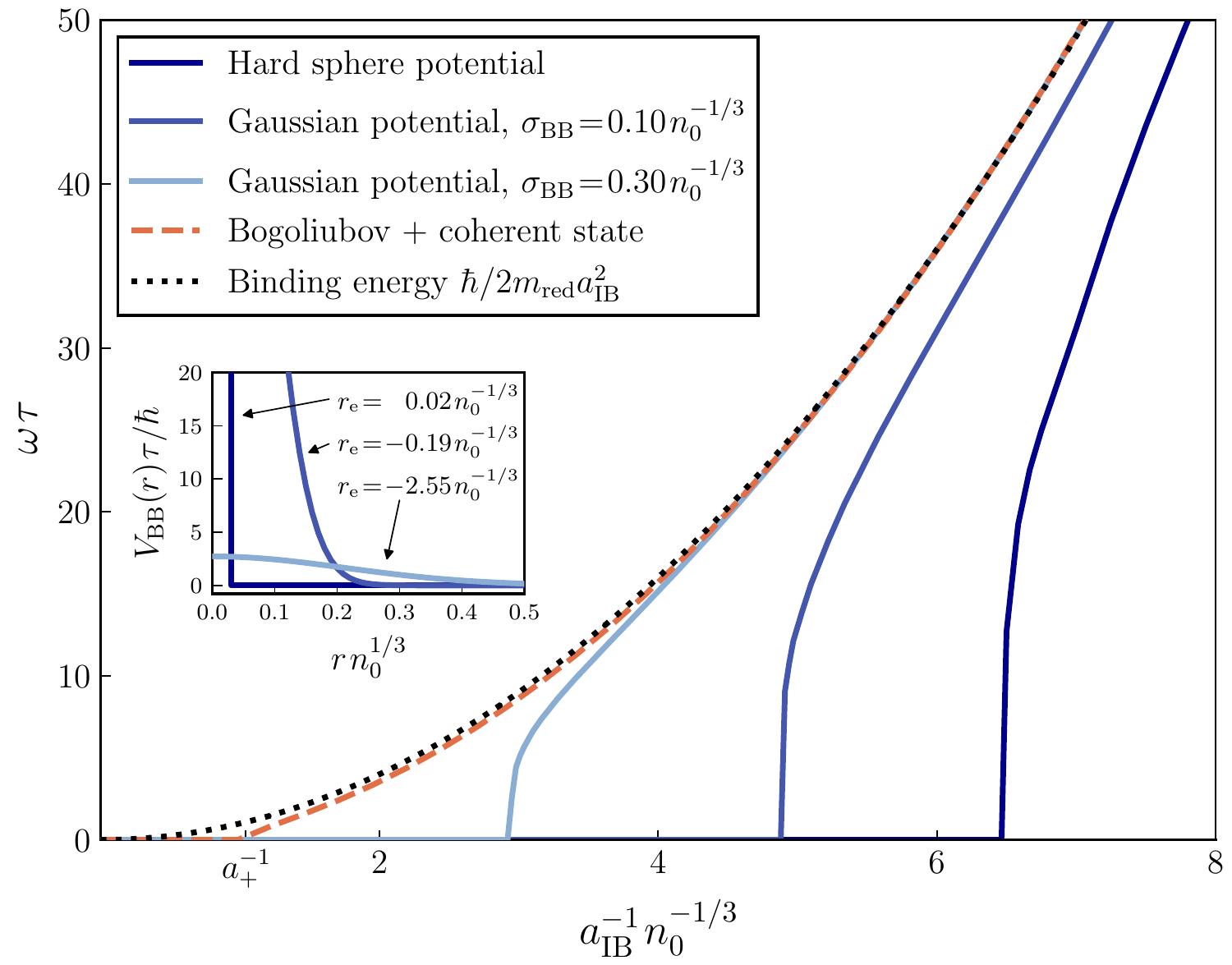}
  \caption{Frequency of coherent oscillations vs.\ impurity-boson
    coupling strength, for different BB potentials with the same
    scattering length $\aBB = 0.03\, n_0^{-1/3}$ but different shapes
    and effective range $r_{\text e}$ (see inset).  The onset of the
    oscillations marks the transition from an attractive to a
    repulsive polaron.  We find that the transition point depends on
    the shape of the BB potential and is shifted further toward the
    repulsive side for a hard-core interaction (black line); different
    IB ranges yield qualitatively similar results.  Softer BB
    potentials (grey), instead, approach the Bogoliubov result
    (dashed) \cite{Drescher2019} with mean-field shifted resonance
    position $a_+^{-1}$ \cite{Shchadilova2016}.  For weaker coupling,
    the oscillation frequencies approach the two-body binding energy
    $E_B$ (dotted) \cite{Drescher2019}.  Parameters and units as in
    Fig.~\ref{fig:contact}.}
  \label{fig:frequencies}
\end{figure}
In Fig.~\ref{fig:contact}, we show the time evolution of Tan's contact
parameter $C = \absq{4π \lim_{r→0} rϕ(r)}$ \cite{Tan2008, Yan2020},
which quantifies the probability to find bosons near the impurity and
determines the short-distance singularity of the pair correlation
function
$g_\text{IB}^{(2)}(\vec r) = \int_{\vec R} \langle \hat
n_\text{B}(\vec R+\vec r) \hat n_\text{I}(\vec R)\rangle = C/(4\pi
r)^2 + \mathcal O(1/r)$, where
$\hat n_\text{I}(\vec x) = \delta(\vec x-\hat{\vec x}_\text{I})$ and
$\hat n_\text{B}(\vec x) = \sum_i \delta(\vec x-\hat{\vec
  x}_{\text{B},i})$ denote the impurity and boson density operators,
resp.  For an ideal BEC, the transition between convergent and
oscillatory behavior is exactly at the resonance $\aIB^{-1} = 0$ and
the frequency is given by the binding energy $E_B$ of one boson to the
impurity \cite{Drescher2019}.  In the presence of BB interactions, the
boundary is shifted towards the repulsive side and convergence
persists at large positive scattering lengths, as shown in
Fig.~\ref{fig:contact}(a).  In the transition region, the system
performs only a finite number of oscillations before assuming
convergent behavior (Fig.~\ref{fig:contact}(b)). This critical region
is narrow and the number of oscillations grows rapidly as the
scattering length is tuned further to the repulsive side
(Fig.~\ref{fig:contact}(c)).

The dynamical transition in Fig.~\ref{fig:contact}(b) is clearly
visible also in the density profiles of the BEC around the impurity
(Fig.~\ref{fig:density}): for short times the condensate density is
strongly depleted in a halo around the impurity that oscillates in
space and time, typical of a repulsive polaron.  For later times,
instead, the depletion propagates outwards and the density profiles of
the BEC attain the shape of an attractive polaron with typically
  $\Delta N=10\dotsc100$ extra bosons in the dressing cloud of the
  impurity \cite{Drescher2019}.  This dynamical
transition regime was inaccessible with previous approaches based on
the Bogoliubov approximation and is the second main result of this
work.

\textit{Position of the resonance in medium and dependence on
  potential shape.---}%
For competing short-range interactions, the question arises whether
the BB interaction is characterized by its scattering length alone, as
in GP and Bogoliubov theory, or whether further details of the
potential need to be taken into account.  To answer this question, we
compute the oscillation frequency after a quench close to the
resonance for different BB potential shapes with the same scattering
length but different height and width, as shown in
Fig.~\ref{fig:frequencies} (inset).  For weak IB coupling
$\aIB ≪ n_0^{-1/3}$ the curves converge, demonstrating that the BB
interaction is indeed fully characterized by the scattering length
alone.  Closer to the resonance, however, the curves differ
significantly.  In particular, the transition point between attractive
and repulsive polaron that marks the onset of oscillations is shifted
much further into the repulsive regime for a hard-sphere potential
than for one that is soft and long-range.  The result obtained from
Bogoliubov theory with a coherent-state ansatz \cite{Drescher2019} is
recovered in the limit of soft potentials where the Born approximation
is valid.  In this case, the transition occurs at the mean-field
shifted resonance position $\aIB = a_+$ \cite{Shchadilova2016,
  Grusdt2017a, Drescher2019} \footnote{The mean-field shift of the
  resonance position is given by
  $a_+^{-1} = (2/π) ∫_0^{∞} d k \, [1 - (\mI + \mB)/(γ_k^2 \mI + γ_k
  \mB)]$, where $γ_k = \sqrt{1 + 16π \aBB n_0/k^2}$.} as indicated in
Fig.~\ref{fig:frequencies}.

To conclude, we have developed a nonlocal extension \eqref{eq:energy
  functional} of Gross-Pitaevskii theory that is able to treat the
impurity-BEC problem at large coupling strengths $\aIB ∼ n₀^{-1/3}$
also close to a Feshbach resonance.  We find a new dynamical
transition region between attractive and repulsive polarons, where the
BEC density profile performs a finite number of oscillations near the
impurity before converging to an attractive polaron profile.  These
phenomena might be observed with current experiments in ultracold
atomic gases \cite{Hu2016, Jorgensen2016, Camargo2018, Yan2020} in
combination with ultrafast interferometry \cite{Cetina2016}, which can
resolve oscillation periods of $100\dotsc1000\mu$s estimated for
lithium-cesium mixtures.  The medium interaction shifts the
interesting transition region further away from the resonance toward
the repulsive side, and this makes it even easier to observe before
three-body recombination occurs on typical timescales of several
milliseconds \cite{Yan2020}.  Our approach captures the medium
backreaction, which could not be described by GPT or Bogoliubov theory
with coherent variational states, and therefore applies also to
ultracold atomic gases with mobile impurities.  In particular,
generalizations to a larger number of mobile impurities, where the
impurity statistics start to matter, will describe also induced
interactions between impurities mediated by the BEC
\cite{Camacho2018}.  More generally, we expect our methods to apply in
systems with bound states and a linear excitation spectrum, where they
provide a generic framework for impurity dressing in medium.

\textit{Note added.}---Recently, a complementary theoretical work
appeared that employs GPT for a finite-range IB potential to study
static Bose polaron properties \cite{Guenther2020}.

\begin{acknowledgments}
  We thank N.~Defenu, A.~Imamoglu, R.~Schmidt, and M.~W. Zwierlein for
  stimulating discussions.  This work is supported by Deutsche
  Forschungsgemeinschaft (DFG, German Research Foundation) via
  Collaborative Research Centre ``SFB1225'' (ISOQUANT) and under
  Germany’s Excellence Strategy ``EXC-2181/1-390900948'' (the
  Heidelberg STRUCTURES Excellence Cluster).
\end{acknowledgments}

\bibliography{references}

\end{document}